\documentclass[aps,prc,preprint,superscriptaddress]{revtex4}

\usepackage{graphicx}
\usepackage{wrapfig}
\usepackage{amsmath,amssymb}

\newcommand{\bra}[1]{\langle {#1} |}
\newcommand{\ket}[1]{| {#1} \rangle}
\newcommand{\inproduct}[2]{\langle #1 | #2 \rangle}
\newcommand{\ainproduct}[2]{\langle\langle #1 | #2 \rangle\rangle}

\begin{document}


\title{
Canonical-basis time-dependent Hartree-Fock-Bogoliubov theory
and linear-response calculations%
}


\author{Shuichiro~Ebata}
\affiliation{RIKEN Nishina Center, Wako-shi, 351-0198, Japan}
\affiliation{Graduate School of Pure and Applied Sciences,
University of Tsukuba, Tsukuba 305-8571, Japan}
\author{Takashi~Nakatsukasa}
\affiliation{RIKEN Nishina Center, Wako-shi, 351-0198, Japan}
\affiliation{Center for Computational Sciences, University of Tsukuba,
Tsukuba 305-8571, Japan}
\author{Tsunenori~Inakura}
\affiliation{RIKEN Nishina Center, Wako-shi, 351-0198, Japan}
\author{Kenichi~Yoshida}
\affiliation{RIKEN Nishina Center, Wako-shi, 351-0198, Japan}
\author{Yukio Hashimoto}
\affiliation{Graduate School of Pure and Applied Sciences,
University of Tsukuba, Tsukuba 305-8571, Japan}
\affiliation{Center for Computational Sciences, University of Tsukuba,
Tsukuba 305-8571, Japan}
\author{Kazuhiro~Yabana}
\affiliation{RIKEN Nishina Center, Wako-shi, 351-0198, Japan}
\affiliation{Graduate School of Pure and Applied Sciences,
University of Tsukuba, Tsukuba 305-8571, Japan}
\affiliation{Center for Computational Sciences, University of Tsukuba,
Tsukuba 305-8571, Japan}


\date{\today}

\begin{abstract}
We present simple equations for a canonical-basis formulation
of the time-dependent Hartree-Fock-Bogoliubov (TDHFB) theory.
The equations are obtained from the TDHFB theory with an approximation
that the pair potential is assumed to be diagonal in the canonical basis.
The canonical-basis formulation significantly reduces the computational cost.
We apply the method to linear-response calculations for even-even light nuclei
and demonstrate its capability and accuracy
by comparing our results with recent calculations
of the quasi-particle random-phase approximation with Skyrme functionals.
We show systematic studies of $E1$ strength distributions
for Ne and Mg isotopes.
The evolution of the low-lying pygmy strength seems to be determined by the
interplay of several factors, including the neutron excess, separation energy,
neutron shell effects, deformation, and pairing.
\end{abstract}


\maketitle


\section{Introduction}
\label{sec:introduction}

The time-dependent Hartree-Fock (TDHF) theory was extensively applied to
studies of nuclear collective phenomena
as a microscopic approach to nuclear dynamics\cite{Neg82}.
Recently, it has been revisited with modern energy density functionals
and more accurate description of nuclear properties
has been achieved\cite{NY05,Mar05,UO06,UO07,WL08,CC09}.
The TDHF theory uses only occupied
orbitals, number of which is equal to the number of particles ($N$),
to describe a variety of nuclear dynamics 
such as heavy-ion scattering,
fusion/fission phenomena, and linear response functions.
However, it neglects the residual interactions in
particle-particle and hole-hole channels,
which becomes problematic especially for open-shell heavy nuclei.
An alternative approach including the pairing correlations
is the time-dependent Hartree-Fock-Bogoliubov (TDHFB) theory\cite{BR86}.
The TDHFB equation is formulated in a similar manner to the TDHF,
however it uses the quasi-particle orbitals instead of the occupied orbitals.
Since the number of the quasi-particle orbitals is, in principle, infinite,
the accurate calculation of TDHFB is presently impractical.
Only recently, a few attempts of the TDHFB calculation have been
performed, but either with a small model space\cite{HN07}
or with restriction to spherical symmetry\cite{ASC08}.

A much simpler approach was proposed by B{\l}ocki and Flocard 
in Ref.~\cite{BF76}.
They gave equations of motion for time-dependent canonical states
$\ket{\phi_k(t)}$ ($k=1,\cdots,M$)
and those for the time-dependent BCS factors $(u_k(t),v_k(t))$.
Since the number of canonical basis is larger than the
particle number but not significantly different ($M\sim N$),
the necessary computational task is roughly the same as that of TDHF.
The similar methods have been applied to studies of heavy-ion reactions
with use of simple functionals\cite{Neg78,CM79,CMS80}.
However, it has never been tested with realistic modern functionals so far,
and we do not know how reliable this approximated scheme is.
In addition, although the equations of motion were provided
for a very schematic pairing functional in Ref.\cite{BF76},
its theoretical foundation seems rather obscure to us.

In this paper, we derive the equations of motion for general functionals
and clarify the approximations/assumptions we need to make.
We call those equations ``Canonical-basis TDHFB'' (Cb-TDHFB) equations.
We apply the method to the linear-response calculations using
the full Skyrme functionals.
The results will be compared with recent calculations of
the quasi-particle random-phase approximation (QRPA),
then demonstrate its feasibility and accuracy.

The paper is organized as follows.
In Sec.~\ref{sec:formalism}, we present the basic equations of the present
method and their derivation.
Especially, we would like to clarify what kind of assumption/approximation
is necessary to justify the Cb-TDHFB equations.
In Sec.~\ref{sec: properties}, we show properties of the Cb-TDHFB,
including gauge invariance, conservation laws, and the small-amplitude limit.
In numerical calculations in this paper,
we adopt a schematic choice for the pairing energy functional,
similar to Ref. \cite{BF76},
which violates the gauge invariance.
In Sec.~\ref{sec: simple_pairing},
we show effect of the violation of the gauge invariance
can be minimized by a special choice
of the gauge condition.
In Sec.~\ref{sec: details}, details of our numerical installation is given.
Then, in Sec.~\ref{sec:numerical_results},
we present numerical results of the real-time calculations of the
linear response 
and compare them with recent QRPA/RPA calculations.
Finally, the conclusion is given in Sec.~\ref{sec:summary}.

\section{Derivation of basic equations}
\label{sec:formalism}

In this section, we derive the basic equations of Cb-TDHFB method.
Using the time-dependent variational principle,
the similar equations were derived by B{\l}ocki and Flocard\cite{BF76}.
However, it was not clear that
what kind of approximation was introduced and how they are different
from the full TDHFB.
We present here a sufficient condition to reduce the TDHFB equations
to those in a simple canonical form.

We start from the density-matrix equation of the TDHFB and
find equations for the canonical-basis states and their occupation-
and pair-probability factors.
In order to clarify our heuristic strategy,
let us start from a simpler case without the pairing correlation.

\subsection{TDHF equation}

The TDHF equation in the density-matrix formalism is written as\cite{RS80}
\begin{equation}
\label{TDHF_density_matrix}
i\frac{\partial}{\partial t} \rho(t) =
\left[ h(t), \rho(t) \right] ,
\end{equation}
where $\rho(t)$ and $h(t)$ are the one-body density operator and
the single-particle (Hartree-Fock) Hamiltonian, respectively.
We now express the one-body density using the time-dependent canonical
single-particle basis,
$\{\ket{\phi_k(t)}\}$, which are assumed to be orthonormal
($\inproduct{\phi_k(t)}{\phi_l(t)}=\delta_{kl}$).
\begin{equation}
\rho(t)=\sum_{k=1}^N \ket{\phi_k(t)}\bra{\phi_k(t)} ,
\end{equation}
where $N$ is the total particle number.
Substituting this into Eq. (\ref{TDHF_density_matrix}),
we have
\begin{equation}
\sum_{k=1}^N \left\{ i\ket{\dot\phi_k(t)}\bra{\phi_k(t)}
+ i\ket{\phi_k(t)}\bra{\dot\phi_k(t)}
\right\}
= \sum_{k=1}^N \left\{
 h(t)\ket{\phi_k(t)}\bra{\phi_k(t)}
- \ket{\phi_k(t)}\bra{\phi_k(t)} h(t)
\right\} .
\end{equation}
the inner product with $\ket{\phi_k(t)}$ leads to
\begin{equation}
\hat{P}
\left( i\frac{\partial}{\partial t} -h(t)\right) \ket{\phi_k(t)} =0
\quad\quad k=1,\cdots,N,
\end{equation}
with $\hat{P}=1-\sum_{k=1}^N \ket{\phi_k(t)}\bra{\phi_k(t)}$.
Here, we used the conservation of the orthonormal property for
the canonical states,
$d/dt \inproduct{\phi_k(t)}{\phi_l(t)}=0$.
This leads to the most general canonical-basis TDHF (Cb-TDHF) equations
\begin{equation}
\label{TDHF_eq}
i\frac{\partial}{\partial t} \ket{\phi_k(t)}
= h(t)\ket{\phi_k(t)} - \sum_{l=1}^N \ket{\phi_l(t)}\eta_{lk}(t),
\quad\quad k=1,\cdots,N,
\end{equation}
where the matrix $\eta_{lk}(t)$ is arbitrary but should be hermitian to
conserve the orthonormal property.
It is easy to see that the time evolution of the density does not
depend on the choice of $\eta_{lk}$.
This is related to the gauge invariance with respect to the
unitary transformations among $\ket{\phi_k(t)}$ ($k=1,\cdots,N$).
The most common choice is $\eta_{lk}=0$, which leads to the
TDHF equation shown in most textbooks.

\subsection{Cb-TDHFB equations}

We now derive Cb-TDHFB equations starting from the generalized
density-matrix formalism.
The TDHFB equation can be written in terms of the generalized density
matrix\cite{BR86} as
\begin{equation}
i\frac{\partial}{\partial t} R = \left[ {\cal H}, R \right] ,
\end{equation}
where
\begin{equation}
R\equiv \begin{pmatrix}
\rho & \kappa \\
-\kappa^* & 1-\rho^* \\
\end{pmatrix}
, \quad\quad
{\cal H}\equiv
\begin{pmatrix}
h & \Delta \\
-\Delta^* & -h^* \\
\end{pmatrix}
.
\end{equation}
This is equivalent to the following equations for
one-body density matrix, $\rho(t)$,
and the pairing-tensor matrix, $\kappa(t)$.
\begin{eqnarray}
\label{TDHFB_1}
i\frac{\partial}{\partial t}\rho(t) &=&
  [h(t),\rho(t)] + \kappa(t) \Delta^*(t) - \Delta(t) \kappa^*(t) ,\\
\label{TDHFB_2}
i\frac{\partial}{\partial t}\kappa(t) &=&
 h(t)\kappa(t)+\kappa(t) h^*(t) + \Delta(t) (1-\rho^*(t)) - \rho(t) \Delta(t) .
\end{eqnarray}

At each instant of time,
we may diagonalize
the density operator $\hat{\rho}$ in
the orthonormal canonical basis, $\{ \phi_k(t), \phi_{\bar k}(t)\}$
with the occupation probabilities $\rho_k$.
For the canonical states, we use the alphabetic indexes such as $k$
for half of the total space indicated by $k>0$.
For each state with $k>0$, there exists a ``paired'' state ${\bar k}<0$
which is orthogonal to all the states with $k>0$.
The set of states $\{ \phi_k, \phi_{\bar k}\}$ generate the whole
single-particle space\footnote{
In the case without pairing ($\Delta=0$),
the canonical pair becomes arbitrary as far as they have the same
occupation probability $\rho_k$ that is either 1 or 0.
}.
We use the Greek letters $\mu,\nu,\cdots$ for indexes of
an adopted representation (complete set) for the single-particle states.
The creation operator of particles
at the state $\ket{\phi_k(t)}$ is expressed as
$\hat{c}_k^\dagger(t) = \sum_\mu \inproduct{\mu}{\phi_k(t)}
 \hat{c}_\mu^\dagger$, and
the TDHFB state is expressed in the canonical (BCS) form as
\begin{equation}
\ket{\Psi(t)}=\prod_{k>0} \left\{
u_k(t) + v_k(t) c_k^\dagger(t) c_{\bar k}^\dagger(t) \right\} \ket{0} .
\end{equation}
For later purposes,
it is convenient to introduce the following notations for two-particle states:
\begin{eqnarray}
\inproduct{\mu\nu}{\phi_k(t)\phi_{\bar k}(t)} &\equiv&
\inproduct{\mu}{\phi_k(t)}
\inproduct{\nu}{\phi_{\bar k}(t)}, \\
\ainproduct{\mu\nu}{\phi_k(t)\phi_{\bar k}(t)} &\equiv&
\inproduct{\mu\nu}{\phi_k(t)\phi_{\bar k}(t)}-
\inproduct{\mu\nu}{\phi_{\bar k}(t)\phi_k(t)} ,
\end{eqnarray}
and for projection operator on a canonical pair of states $(k,{\bar k})$,
\begin{equation}
\hat{\pi}_k(t) \equiv \ket{\phi_k(t)}\bra{\phi_k(t)}
  + \ket{\phi_{\bar k}(t)}\bra{\phi_{\bar k}(t)} .
\end{equation}
Then, it is easy to show the following properties ($k,l>0$):
\begin{eqnarray}
\sum_{\mu\nu}\inproduct{\mu\nu}{\phi_k(t)\phi_{\bar k}(t)}
             \inproduct{\phi_l(t)\phi_{\bar l}(t)}{\mu\nu}
&=&\delta_{kl},  \\
\sum_{\mu\nu}\ainproduct{\mu\nu}{\phi_k(t)\phi_{\bar k}(t)}
             \ainproduct{\phi_l(t)\phi_{\bar l}(t)}{\mu\nu}
&=&2\delta_{kl}, \\
\sum_{\sigma}\ainproduct{\mu\sigma}{\phi_k(t)\phi_{\bar k}(t)}
             \ainproduct{\phi_l(t)\phi_{\bar l}(t)}{\nu\sigma}
&=&\delta_{kl} \bra{\mu}\hat{\pi}(t)_k\ket{\nu} ,\\
\sum_{\sigma}\ainproduct{\mu\sigma}{\phi_k(t)\phi_{\bar k}(t)}
             \bra{\nu} \hat\pi_l \ket{\sigma}
&=&\delta_{kl} \ainproduct{\mu\nu}{\phi_k(t)\phi_{\bar k}(t)} .
\end{eqnarray}
Using these notations,
the density and the pairing-tensor matrixes
are given by
\begin{eqnarray}
\label{rho_mn}
\rho_{\mu\nu}(t)&=& \sum_{k>0}
\rho_k(t) \bra{\mu}\hat{\pi}_k(t)\ket{\nu} , \\
\label{kappa_mn}
\kappa_{\mu\nu}(t) &=& \sum_{k>0} \kappa_k(t)
\ainproduct{\mu\nu}{\phi_k(t)\phi_{\bar k}(t)} ,
\end{eqnarray}
where $\rho_k(t)$ and $\kappa_k(t)$ are occupation and pair probabilities,
respectively.
In terms of the BCS factors of $(u,v)$\cite{RS80}, they are given as
$\rho_k(t)=|v_k(t)|^2$ and $\kappa_k(t)=u_k^*(t) v_k(t)$.
It should be noted that the canonical pair of states,
$\ket{\phi_k(t)}$ and $\ket{\phi_{\bar k}(t)}$,
are assumed to be orthonormal
but not necessarily related with each other by the time reversal,
$\ket{\phi_{\bar k}} \neq T\ket{\phi_k}$.

Thanks to the orthonormal property, we can invert Eqs. (\ref{rho_mn}) and
(\ref{kappa_mn}) for $\rho_k$ and $\kappa_k$,
\begin{eqnarray}
\label{rho_k}
\rho_k(t)&=&
 \sum_{\mu\nu}
 \inproduct{\phi_k(t)}{\mu}\rho_{\mu\nu}(t)\inproduct{\nu}{\phi_k(t)}
 =\sum_{\mu\nu}
 \inproduct{\phi_{\bar k}(t)}{\mu}\rho_{\mu\nu}(t)
  \inproduct{\nu}{\phi_{\bar k}(t)} , \\
\label{kappa_k}
\kappa_k(t)&=&\sum_{\mu\nu} 
\inproduct{\phi_k(t)\phi_{\bar k}(t)}{\mu\nu} \kappa_{\mu\nu}(t) 
=\frac{1}{2} \sum_{\mu\nu}
\ainproduct{\phi_k(t)\phi_{\bar k}(t)}{\mu\nu} \kappa_{\mu\nu}(t) .
\end{eqnarray}
With help of Eq. (\ref{rho_mn}),
the derivative of $\rho_k(t)$ with respect to time $t$ leads to
\begin{eqnarray}
i\frac{d}{dt}\rho_k(t) &=&\sum_{\mu\nu}
\inproduct{\phi_k(t)}{\mu}i\frac{d\rho_{\mu\nu}}{dt}\inproduct{\nu}{\phi_k(t)}
+i\rho_k(t) \frac{d}{dt} \inproduct{\phi_k(t)}{\phi_k(t)} \nonumber\\
&=&\sum_{\mu\nu}
\inproduct{\phi_k(t)}{\mu}i\frac{d\rho_{\mu\nu}}{dt}\inproduct{\nu}{\phi_k(t)}
\nonumber \\
&=&\sum_{\mu\nu} \left\{
\kappa_k(t) 
\Delta_{\mu\nu}^*(t)\inproduct{\nu\mu}{\phi_k(t)\phi_{\bar k}(t)}
+
 \kappa_k^*(t)
\Delta_{\mu\nu}(t) \inproduct{\phi_k(t)\phi_{\bar k}(t)}{\mu\nu}
\right\} .
\end{eqnarray}
We used the assumption of norm conservation for the second equation, and
used the TDHFB equation (\ref{TDHFB_1}) in the last equation.
Since the pair potential $\Delta_{\mu\nu}(t)$
is anti-symmetric, it is written in a simple form as
\begin{eqnarray}
\label{drho_dt}
&&
i\frac{d}{dt}\rho_k(t) =
\kappa_k(t) \Delta_k^*(t)
-\kappa_k^*(t) \Delta_k(t) , \\
\label{Delta_k}
&&\Delta_k(t) \equiv -\sum_{\mu\nu}
\Delta_{\mu\nu}(t)
 \inproduct{\phi_k(t)\phi_{\bar k}(t)}{\mu\nu}
=-\frac{1}{2} \sum_{\mu\nu}
\Delta_{\mu\nu}(t)
 \ainproduct{\phi_k(t)\phi_{\bar k}(t)}{\mu\nu} .
\end{eqnarray}
In case that the pair potential is computed from a two-body
interaction $v$ as
$\Delta_{\mu\nu}(t)=\sum_{\alpha\beta}
v_{\mu\nu,\alpha\beta} \kappa_{\alpha\beta}(t)$,
the gap parameters, $\Delta_k(t)$, are identical
to those of the BCS approximation \cite{RS80}.
\begin{equation}
\Delta_k(t)=-\sum_{l>0} \kappa_l(t) (v_{k\bar{k},l\bar{l}} - 
                                v_{k\bar{k},\bar{l}l})
\equiv -\sum_{l>0} \kappa_l(t) {\bar v}_{k\bar{k},l\bar{l}} .
\end{equation}
It should be noted here that the two-body 
matrix elements $v_{k\bar{k},l\bar{l}}$ 
(and the anti-symmetric ${\bar v}_{k\bar{k},l\bar{l}}$) are
time-dependent because the canonical basis,
$(k,{\bar k})$ and $(l,{\bar l})$, are time-dependent.

In the same way, we evaluate the time derivative of $\kappa_k(t)$ as
\begin{equation}
i\frac{d}{dt}\kappa_k(t)=\sum_{\mu\nu}
\inproduct{\phi_k(t)\phi_{\bar k}(t)}{\mu\nu}
i\frac{d\kappa_{\mu\nu}}{dt}
+i\kappa_k(t) \left( \inproduct{\frac{d\phi_k}{dt}}{\phi_k(t)}
                   +\inproduct{\frac{d\phi_{\bar k}}{dt}}{\phi_{\bar k}(t)}
\right) .
\end{equation}
Then, using the TDHFB equation (\ref{TDHFB_2}), we obtain
\begin{equation}
\label{dkappa_dt}
i\frac{d}{dt}\kappa_k(t) =
\left(
\eta_k(t)+\eta_{\bar k}(t)
\right) \kappa_k(t) +
\Delta_k(t) \left( 2\rho_k(t) -1 \right) ,
\end{equation}
where $\eta_k(t)\equiv\bra{\phi_k(t)}h(t)\ket{\phi_k(t)}
+i\inproduct{\frac{\partial\phi_k}{\partial t}}{\phi_k(t)}$.

The time-dependent equations for $\rho_k(t)$ and $\kappa_k(t)$
are now given in rather simple forms
as Eqs. (\ref{drho_dt}) and (\ref{dkappa_dt}).
So far, their derivation is solely based on the TDHFB equations,
utilizing the fact that
$\rho_{\mu\nu}(t)$ and $\kappa_{\mu\nu}(t)$
can be expressed by the orthonormal canonical basis,
$\ket{\phi_k(t)}$ and $\ket{\phi_{\bar k}(t)}$,
and their occupation and pair probabilities,
$\rho_k(t)$ and $\kappa_k(t)$.
However, in general, the time evolution of the canonical basis is
not given by a simple equation.
Therefore, we now introduce an assumption (approximation) that
the pair potential is written as
\begin{equation}
\label{Delta_mn}
\Delta_{\mu\nu}(t) = -\sum_{k>0}
\Delta_k(t) \ainproduct{\mu\nu}{\phi_k(t)\phi_{\bar k}(t)} .
\end{equation}
This satisfies Eq. (\ref{Delta_k}),
but in general,
Eq. (\ref{Delta_k}) can not be inverted because
the two-particle states $\ket{\phi_k\phi_{\bar k}}$
do not span the whole space.
In other words, we only take into account the pair potential
of the ``diagonal'' parts in the canonical basis,
$\Delta_{k{\bar l}}=-\Delta_k\delta_{kl}$.
In the stationary limit ($\ket{\phi_{\bar k}}=T\ket{\phi_k}$),
this is equivalent to the ordinary BCS approximation
(see Sec.~\ref{sec: static_limit}).

With the approximation of Eq. (\ref{Delta_mn}),
it is easy to see that
the TDHFB equations, (\ref{TDHFB_1}) and (\ref{TDHFB_2}),
are consistent with the following equations:
\begin{equation}
\label{dphi_dt}
i\frac{\partial}{\partial t} \ket{\phi_k(t)} =
(h(t)-\eta_k(t))\ket{\phi_k(t)} , \quad\quad
i\frac{\partial}{\partial t} \ket{\phi_{\bar k}(t)} =
(h(t)-\eta_{\bar k}(t))\ket{\phi_{\bar k}(t)} .
\end{equation}

In summary,
the Cb-TDHFB equations consists of Eqs.
(\ref{dphi_dt}), 
(\ref{drho_dt}), and (\ref{dkappa_dt}).
To derive these equations from the TDHFB equations,
we have assumed the diagonal property of the pair
potential, Eq. (\ref{Delta_mn}).

\section{Properties of the Cb-TDHFB equations}
\label{sec: properties}

\subsection{Gauge invariance}

The $\eta_k(t)$ and $\eta_{\bar k}(t)$,
in Eqs. (\ref{dkappa_dt}) and (\ref{dphi_dt}),
must be real to conserve the orthonormal
property, however, they are arbitrary.
This is related to the phase degrees of freedom of the canonical
states.
The Cb-TDHFB equations, (\ref{drho_dt}),
(\ref{dkappa_dt}) and (\ref{dphi_dt}), are invariant
with respect to the following gauge transformations
with arbitrary real functions, $\theta_k(t)$ and $\theta_{\bar k}(t)$.
\begin{eqnarray}
\label{gauge_transf_1}
\ket{\phi_k}\rightarrow e^{i\theta_k(t)}\ket{\phi_k}
\quad &\mbox{and}& \quad
\ket{\phi_{\bar k}}\rightarrow e^{i\theta_{\bar k}(t)}\ket{\phi_{\bar k}}
\\
\label{gauge_transf_2}
\kappa_k\rightarrow e^{-i(\theta_k(t)+\theta_{\bar k}(t))}\kappa_k
\quad &\mbox{and}& \quad
\Delta_k\rightarrow e^{-i(\theta_k(t)+\theta_{\bar k}(t))}\Delta_k
\end{eqnarray}
simultaneously with 
$$
\eta_k(t)\rightarrow \eta_k(t)+\frac{d\theta_k}{dt}
\quad \mbox{and} \quad
\eta_{\bar k}(t)\rightarrow \eta_{\bar k}(t)+\frac{d\theta_{\bar k}}{dt} .
$$
The phase relations of Eq. (\ref{gauge_transf_2}) are obtained from
Eqs. (\ref{kappa_k}) and (\ref{Delta_k}).

\subsection{Conservation laws}

\subsubsection{Orthonormality of canonical states}

Apparently, Eq. (\ref{dphi_dt}) conserves the orthonormal property
of canonical states, as far as $\eta_k$ are real.
\begin{equation}
i\frac{\partial}{\partial t} \inproduct{\phi_k(t)}{\phi_l(t)}
= \bra{\phi_k(t)} \{(h(t)-\eta_l(t))-(h^\dagger(t)-\eta_k(t)) \}
\ket{\phi_l(t)} = 0 .
\end{equation}
Here, we assume $\inproduct{\phi_k(t)}{\phi_l(t)}=\delta_{kl}$
at time $t$.

\subsubsection{Average particle number}

The average particle number also conserves because
\begin{equation}
i\frac{d}{dt} N(t) = 2i\frac{d}{dt} \sum_{k>0} \rho_k(t)
= 2\sum_{k>0} (\kappa_k(t)\Delta_k^*(t) - \kappa_k^*(t) \Delta_k(t) )
= 0 ,
\end{equation}
where we used the expression of the pairing energy,
Eq. (\ref{E_pair}), for the last equation.

\subsubsection{Average total energy}

Time variation of the energy functional $E[\rho,\kappa]$ can be divided
into two: $dE/dt=dE/dt|_\rho + dE/dt|_\kappa$.
The variation of energy associated with the normal-density fluctuation is
\begin{equation}
\label{dE_drho}
i\left.\frac{dE}{dt}\right|_\rho=
i\sum_{\mu\nu}
\frac{\partial E}{\partial \rho_{\mu\nu}} \frac{d\rho_{\mu\nu}}{dt}
= i\sum_{k>0} \frac{d\rho_k}{dt} (\epsilon_k(t)+\epsilon_{\bar k}(t)) ,
\end{equation}
where $\epsilon_k(t)=\bra{\phi_k(t)} h(t) \ket{\phi_k(t)}$.
This equation has an intuitive physical interpretation.
The energy carried by a canonical state $\ket{\phi_k}$
is $\epsilon_k(t)\times\rho_k$.
If the occupation probability is fixed during the
time evolution, 
the right-hand side of Eq. (\ref{dE_drho}) vanishes.
This corresponds to the cases such as the TDHF and its extension with fixed
BCS occupation probabilities.
In the TDHFB, the energy variation in Eq. (\ref{dE_drho}) transfers
from/to the pairing energy.
In fact, time variation due to the pairing tensors-produce fluctuation
produces
\begin{equation}
\label{dE_dkappa}
i\left.\frac{dE}{dt}\right|_\kappa=
i\frac{1}{2}\sum_{\mu\nu} \left(
\frac{\partial E}{\partial \kappa_{\mu\nu}} \frac{d\kappa_{\mu\nu}}{dt}
+\frac{\partial E}{\partial \kappa_{\mu\nu}^*} \frac{d\kappa_{\mu\nu}^*}{dt}
\right)
= \sum_{k>0} (\kappa_k^* \Delta_k - \kappa_k \Delta_k^*)
 (\epsilon_k(t)+\epsilon_{\bar k}(t)) ,
\end{equation}
where Eq. (\ref{Delta_mn}) is used.
Because of Eq. (\ref{drho_dt}), two contributions of
Eqs. (\ref{dE_drho}) and (\ref{dE_dkappa}) always cancel and
the total energy is conserved.
This is natural because the Cb-TDHFB equations satisfy the TDHFB equations,
(\ref{TDHFB_1}) and (\ref{TDHFB_2}), for which the conservation of the total
energy in TDHFB is well known\cite{BR86}.

\subsection{Stationary solution}
\label{sec: static_limit}

When we assume that all the canonical states are eigenstates of
the time-independent single-particle Hamiltonian $h_0$.
\begin{eqnarray}
&&
\ket{\phi_k(t)}=\ket{\phi_k^0} e^{i\theta_k(t)} , \quad
\ket{\phi_{\bar k}(t)}=\ket{\phi_{\bar k}^0} e^{i\theta_{\bar k}(t)} , \\
\label{phi_0}
&&
h_0\ket{\phi_k^0} = \epsilon_k^0 \ket{\phi_k^0}, \quad
h_0\ket{\phi_{\bar k}^0} = \epsilon_k^0 \ket{\phi_{\bar k}^0},
\end{eqnarray}
where $\ket{\phi_{\bar k}}=T\ket{\phi_k}$ have the same eigenvalues
$\epsilon_k^0$ as $\ket{\phi_k}$.
Here, $d\theta_k/dt=i\inproduct{\partial\phi_k/\partial t}{\phi_k}$ and
$d\theta_{\bar k}/dt=
i\inproduct{\partial\phi_{\bar k}/\partial t}{\phi_{\bar k}}$
are arbitrary real functions of $t$.
$\kappa_k(t)$ and $\Delta_k(t)$ should have a common time-dependent phase
associated with the chemical potential $\lambda$ as $e^{-2i\lambda t}$.
In addition to this,
according to their definition, Eqs. (\ref{kappa_k}) and (\ref{Delta_k}),
they have the following additional phases connected with the phases
of the canonical states.
\begin{eqnarray}
\label{kappa_k_static}
\kappa_k(t)&=&\kappa_k^0 \exp\{-i(2\lambda t+\theta_k(t)+\theta_{\bar k}(t))\} ,\\
\label{Delta_k_static}
\Delta_k(t)&=&\Delta_k^0 \exp\{-i(2\lambda t+\theta_k(t)+\theta_{\bar k}(t))\} ,
\end{eqnarray}
The stationary case of Eq. (\ref{drho_dt}), $d\rho_k^0/dt=0$, indicates
that $\kappa_k^0$ and $\Delta_k^0$ have the same arguments to make
$\kappa_k(t) \Delta_k^*(t)$ real.
If all the pairing matrix elements are real,
we can choose both $\kappa_k^0$ and $\Delta_k^0$ are real.
Then, Eq. (\ref{dkappa_dt}) reduces to
\begin{equation}
\label{kappa_0}
2(\epsilon_k^0 - \lambda) \kappa_k^0 + \Delta_k^0 (2\rho_k^0 -1) = 0 .
\end{equation}
This is consistent with the ordinary BCS result.
\begin{eqnarray}
\rho_k^0 &=& \frac{1}{2} \left( 1 -
\frac{\epsilon_k^0 -\lambda}{\sqrt{(\epsilon_k^0-\lambda)^2+(\Delta_k^0)^2}}
\right) , \\
\kappa_k^0 &=& \frac{1}{2}
\frac{\Delta_k^0}{\sqrt{(\epsilon_k^0-\lambda)^2+(\Delta_k^0)^2}} .
\end{eqnarray}

\subsection{Small amplitude limit and the Nambu-Goldstone modes}

It is known that the small-amplitude approximation for the TDHFB around the
HFB ground state is identical to the QRPA.
In the QRPA, when the ground state (HFB state) breaks continuous
symmetry of the Hamiltonian, the Nambu-Goldstone modes appear as
the zero-energy modes.
In this section, we show that this is also true for the small-amplitude limit
of the Cb-TDHFB.

The ground state is given by $\ket{\phi_k^0}$, $\ket{\phi_{\bar k}^0}$,
$\kappa_k^0$, and $\rho_k^0$
which satisfy Eqs. (\ref{phi_0}) and (\ref{kappa_0}).
Extracting trivial phase factors,
$\xi_k(t)\equiv \int_0^t \left\{ \eta_k(t')-\epsilon_k^0 \right\} dt'$,
we express the time-dependent quantities as follows:
\begin{eqnarray}
&&
\ket{\phi_k(t)}=\left\{ \ket{\phi_k^0} + \ket{\delta\phi_k(t)}\right\}
                 e^{i\xi_k(t)} , \quad
\ket{\phi_{\bar k}(t)}=\left\{ \ket{\phi_{\bar k}^0} +
    \ket{\delta\phi_{\bar k}(t)}\right\} e^{i\xi_{\bar k}(t)} , \\
&&
\kappa_k(t)=\left\{ \kappa_k^0 + \delta\kappa_k(t) \right\}
                 e^{-i\{\xi_k(t)+\xi_{\bar k}(t)+2\lambda t\}} , \quad
\label{Delta_phase}
\Delta_k(t)=\left\{ \Delta_k^0 + \delta\Delta_k(t) \right\}
                 e^{-i\{\xi_k(t)+\xi_{\bar k}(t)+2\lambda t\}} , \\
&&
\rho_k(t)=\rho_k^0 + \delta\rho_k(t) , \quad
h(t)=h_0 + \delta h(t) ,
\end{eqnarray}
Substituting these
into Eqs. (\ref{dphi_dt}), (\ref{drho_dt}), and (\ref{dkappa_dt}),
they lead to the following equations in the linear order with respect to
the fluctuation.
\begin{eqnarray}
\label{ddeltaphi_dt}
i\frac{\partial}{\partial t} \ket{\delta\phi_k(t)}
&=& (h_0 - \epsilon_k^0) \ket{\delta\phi_k(t)}+\delta h(t)\ket{\phi_k^0} ,
\quad (k\leftrightarrow \bar k) , \\
\label{ddeltarho_dt}
i\frac{\partial}{\partial t} \delta\rho_k(t) &=&
\Delta_k^{0*} \delta\kappa_k(t) + \kappa_k^0 \delta\Delta_k^*(t)
 - \mbox{ c.c.} ,\\
\label{ddeltakappa_dt}
i\frac{\partial}{\partial t} \delta\kappa_k(t) &=&
2(\epsilon_k^0-\lambda)\delta\kappa_k(t)
+(2\rho_k^0-1) \delta\Delta_k(t)
+2\Delta_k^0\delta\rho_k(t) .
\end{eqnarray}
When these fluctuating parts have specific oscillating frequency $\omega$,
they correspond to the normal modes.
The zero-energy modes correspond to stationary normal-mode solutions
with $\omega=0$.

\subsubsection{Translation and rotation}
\label{sec: translation_rotation}

When the HFB ground state spontaneously violates the translational
(rotational) symmetry, there are generators, $\vec{P}$ ($\vec{J}$),
which transform the ground state into a new state
but keep the energy invariant.
Here, let us denote one of those hermitian generators, $S$.
The transformation with respect to the generator $S$
with real parameter $\alpha$ leads to
\begin{eqnarray}
\ket{\phi_k^0} &\rightarrow& \ket{\phi_k^0(\alpha)} =
e^{i\alpha S} \ket{\phi_k^0}
\quad (k \leftrightarrow \bar k) , \\
h_0 &\rightarrow&
h_0(\alpha) = e^{i\alpha S} h_0 e^{-i\alpha S} ,
\end{eqnarray}
with $\rho_k(\alpha)=\rho_k^0$, $\kappa_k(\alpha)=\kappa_k^0$,
$\epsilon_k(\alpha)=\epsilon_k^0$, and
$\Delta_k(\alpha)=\Delta_k^0$.
These transformed quantities should also satisfy Eq. (\ref{phi_0}).
\begin{equation}
(h_0(\alpha) - \epsilon_k^0) \ket{\phi_k^0(\alpha)} = 0 .
\end{equation}
In the linear order with respect to the parameter $\alpha$, we have
\begin{equation}
\label{NG_1}
i\alpha (h_0 - \epsilon_k^0) S \ket{\phi_k^0} 
+ i\alpha [S,h_0] \ket{\phi_k^0} = 0 .
\end{equation}
Equation (\ref{NG_1}) means that
$\ket{\delta\phi_k^S}\equiv i \alpha S \ket{\phi_k^0}$ and
$\delta h_S\equiv i\alpha [S,h_0]$
correspond to a normal-mode solution with $\omega=0$
for Eq. (\ref{ddeltaphi_dt}).
$\delta\rho_k^S=0$, $\delta\kappa_k^S=0$, and
$\delta\Delta_k^S=0$
also satisfy Eqs. (\ref{ddeltarho_dt})
and (\ref{ddeltakappa_dt}).
Therefore, the Nambu-Goldstone modes related to the spontaneous breaking of
the translational and rotational symmetries become zero-energy modes
in the small-amplitude Cb-TDHFB equations.

\subsubsection{Pairing rotation}

When the ground state is in the superfluid phase,
we have $\kappa_k^0\neq 0$ at least for a certain $k$.
The ground state can be transformed into a new state
by operation of $e^{i\theta N}$ 
where $N$ is the number operator.
This transformation changes the phase of $\kappa_k$ and $\Delta_k$ but keeps
the other quantities invariant.
\begin{eqnarray}
\delta\kappa_k^N &=& e^{2i\theta} \kappa_k^0 -\kappa_k^0\approx
2i\theta\kappa_k^0, \quad
\delta\Delta_k^N = e^{2i\theta} \Delta_k^0 -\Delta_k^0\approx
2i\theta\Delta_k^0, \\
\delta\rho_k^N &=& \delta h_N = 0, \quad
\ket{\delta\phi_k^N} = \ket{\delta\phi_{\bar k}^N} = 0 .
\end{eqnarray}
Using Eq. (\ref{kappa_0}), it is easy to see that
these quantities correspond to an $\omega=0$ solution of
Eqs. (\ref{ddeltaphi_dt}), (\ref{ddeltarho_dt}), and (\ref{ddeltakappa_dt}).
Thus, the pairing rotational modes 
appear as the zero energy modes as well.

\subsubsection{Particle-particle (hole-hole) RPA}
\label{sec: pairing_vibration}

The Cb-TDHFB equation (\ref{ddeltaphi_dt}) seems to be independent from
the rest of equations, (\ref{ddeltarho_dt}) and (\ref{ddeltakappa_dt}),
at first sight.
However,
this is not true in general, because $\delta\Delta_k(t)$ depend on
$\ket{\delta\phi_k(t)}$ and $\delta h(t)$ depends on $\delta\rho_k(t)$.
In contrast,
when the ground state is in the normal phase ($\kappa_k^0=\Delta_k^0=0$),
$\delta\Delta_k(t)$ becomes independent from $\ket{\delta\phi_l(t)}$,
and we have $\delta\rho_k(t)=0$.
This means that the particle-hole (p-h) channel is exactly decoupled from
the particle-particle (p-p) and hole-hole (h-h) channels.
It is well-known that TDHF equations in the small-amplitude limit,
(\ref{ddeltaphi_dt}),
reduce to the RPA equation in the ph-channel\cite{RS80,BR86,NIY07}.
Thus, we here discuss properties of the p-p and h-h channels.

The p-p and h-h dynamics are described by the following equations.
\begin{equation}
\label{pair_vib}
i\frac{\partial}{\partial t} \delta\kappa_k(t) =
2\epsilon_k^0 \delta\kappa_k(t)
\pm \delta\Delta_k(t) ,
\end{equation}
where the sign $+$ ($-$) is for hole (particle) orbitals, and
we omit the chemical potential $\lambda$.
For the p-p channel ($\omega=E_{N+2}-E_N$),
a normal mode with frequency $\omega$ is described by
$\delta\kappa_p=X_p e^{-i\omega t}$ for particle orbitals
($|p|>N/2$)
and $\delta\kappa_h=-Y_h e^{-i\omega t}$ for hole orbitals
($|h|\leq N/2$).
For the h-h channel ($\omega=E_{N-2}-E_N$),
it is described by
$\delta\kappa_h=X_h e^{-i\omega t}$ for hole orbitals
($|h|\leq N/2$)
and $\delta\kappa_p=-Y_p e^{-i\omega t}$ for particle orbitals
($|k|> N/2$).
Equation (\ref{pair_vib}) can be rewritten in a matrix form as
\begin{equation}
\begin{pmatrix}
2\epsilon_p^0\delta_{pp'} +{\bar v}_{p{\bar p}p'{\bar p'}}
        & -{\bar v}_{p{\bar p}h'{\bar h'}} \\
-{\bar v}_{h{\bar h}p'{\bar p'}} 
        & -2\epsilon_h^0\delta_{hh'} +{\bar v}_{h{\bar h}h'{\bar h'}}
\end{pmatrix}
\begin{pmatrix}
Z_{p'} \\
Z_{h'}
\end{pmatrix}
= \omega
\begin{pmatrix}
1 & 0 \\
0 & -1 \\
\end{pmatrix}
\begin{pmatrix}
Z_p \\
Z_h
\end{pmatrix} ,
\end{equation}
where
$Z_p= X_p$ ($Z_p= Y_p$) and $Z_h=Y_h$ ($Z_h=X_h$) for the p-p (h-h) channel.
This is equivalent to the p-p and h-h RPA in the BCS approximation\cite{RS80}.

\section{Cb-TDHFB equations with a simple pairing energy functional
and gauge condition}
\label{sec: simple_pairing}

\subsection{Pairing energy functional}

Normally, the pairing energy functional is bi-linear with respect
to $\kappa_{\mu\nu}$ and $\kappa_{\mu\nu}^*$.
For instance, when it is calculated from the two-body interaction,
it is given by
\begin{equation}
E_\kappa(t)
= \sum_{\mu,\nu,\rho,\sigma}
v_{\mu\nu,\rho\sigma} \kappa^*_{\mu\nu}(t)
 \kappa_{\rho\sigma}(t) .
\end{equation}
Thus, the pairing energy can be also written as
\begin{eqnarray}
E_\kappa(t)
&=& \frac{1}{2}\sum_{\mu\nu}
\kappa_{\mu\nu}(t) \Delta_{\mu\nu}^*(t)
= \frac{1}{2}\sum_{\mu\nu}
\kappa_{\mu\nu}^*(t) \Delta_{\mu\nu}(t) \\
\label{E_pair}
&=& -\sum_{k>0}
\kappa_k(t) \Delta_k^*(t)
= -\sum_{k>0}
\kappa_k^*(t) \Delta_k(t) .
\end{eqnarray}

For numerical calculations in the present study,
we adopt a schematic pairing functional in a form of
\begin{equation}
\label{E_G}
E_g(t)=-\sum_{k,l>0} G_{kl} \kappa_k^*(t) \kappa_l(t) ,
=-\sum_{k>0} \kappa_k^*(t) \Delta_k(t) ,
\quad
\Delta_k(t)= \sum_{l>0} G_{kl} \kappa_l(t) ,
\end{equation}
where $G_{kl}$ is a hermitian matrix.
This pairing functional produces a pair potential which
is diagonal in the canonical basis.
This is consistent with the approximation of Eq.~(\ref{Delta_mn}).
However, the functional violates the gauge invariance
(\ref{gauge_transf_2}), because
\begin{equation}
\sum_{l>0} G_{kl} e^{-i\theta_l+\theta_{\bar l}} \kappa_l(t) 
\neq e^{-i\theta_k+\theta_{\bar k}} \sum_{k>0} G_{kl} \kappa_l(t) .
\end{equation}
The violation comes from the fact that the $\Delta_k(t)$ in this
schematic definition no longer hold the correct phase relation to
canonical states $(k,{\bar k})$,
according to the definition of Eq. (\ref{Delta_k}).
Therefore, we require the gauge condition of
$\inproduct{\frac{\partial \phi_k}{\partial t}}{\phi_k} =
\inproduct{\frac{\partial \phi_{\bar k}}{\partial t}}{\phi_{\bar k}} =0$,
so as to minimize the phase change of canonical states.
This means that we choose the gauge parameters $\eta_k(t)$ as
\begin{equation}
\label{gauge_fix}
\eta_k(t)=\epsilon_k(t)=\bra{\phi_k(t)} h(t) \ket{\phi_k(t)}, \quad
\eta_{\bar k}(t)=\epsilon_{\bar k}(t) 
=\bra{\phi_{\bar k}(t)} h(t) \ket{\phi_{\bar k}(t)} .
\end{equation}

\subsection{Properties of Cb-TDHFB equations with $E_g$}
\label{sec: properties_E_G}

The Cb-TDHFB equations with the simple pairing functional (\ref{E_G})
keep the following desired properties, if we adopt the special gauge condition
(\ref{gauge_fix}).
The details are presented in Appendix.
\begin{enumerate}
\item Conservation law
 \begin{enumerate}
 \item Conservation of orthonormal property of the canonical states
 \item Conservation of average particle number
 \item Conservation of average total energy
 \end{enumerate}
\item The stationary solution corresponds to the HF+BCS solution.
\item Small-amplitude limit
 \begin{enumerate}
 \item The Nambu-Goldstone modes are zero-energy normal-mode solutions.
 \item If the ground state is in the normal phase, the equations are
         identical to the particle-hole, particle-particle, and hole-hole
         RPA with the BCS approximation.
 \end{enumerate}
\end{enumerate}
Among these properties, 1(a) and 1(b) do not depend on the choice of
the gauge, however, the other properties are guaranteed only with the
special choice of gauge (\ref{gauge_fix}).

\section{Details of numerical calculations}
\label{sec: details}

\subsection{Treatment of the pairing energy functional}

In numerical calculations, we start from the HF+BCS calculation for the
ground state.
The pairing energy is calculated for the constant monopole pairing
interaction with a smooth truncation for the model space.
We follow the prescription given by Tajima et al\cite{TTO96}, which
is equivalent to the following choice of $G_{kl}$
of Eq. (\ref{E_G}).
\begin{equation}
\label{gff}
G_{kl}=g f(\epsilon_k^0) f(\epsilon_l^0) ,
\end{equation}
with a constant real parameter $g$.
The cut-off function $f(\varepsilon)$,
which depends on the ground-state single-particle energies,
is in the following form
\begin{equation}
f(\varepsilon)=\left( 1+\exp\left[ 
\frac{\varepsilon - \epsilon_{\rm c}}{0.5 \mbox{ MeV}} \right]\
 \right)^{-1/2} \theta (e_{\rm c}-\varepsilon),
\end{equation}
with the cut-off energies 
\begin{eqnarray}
\epsilon_{\rm c} = \tilde\lambda+5.0\ {\rm MeV},\quad
e_{\rm c} = \epsilon_{\rm c}+2.3\ {\rm MeV} ,
\end{eqnarray}
where $\tilde\lambda$ is the average of the highest occupied level
and the lowest unoccupied level in the HF state.
Here, the cut-off parameter $e_{\rm c}$ is necessary to prevent
occupation of spatially unlocalized single-particle states,
known as the problem of unphysical gas near the drip line.
For neutrons, if $e_{\rm c}$ becomes positive, we replace it by zero.

To determine the pairing strength constant $g$ for each nuclei,
we again follow the prescription of Ref. \cite{TTO96} which is
practically identical to the one in Ref. \cite{Bra72}.
For light nuclei ($A<50$), we replace $g$ with 0.6 MeV 
when the calculated value exceeds 0.6 MeV.
The pairing force strengths $G_{kl}$ are
calculated for the ground state and kept constant during the time evolution.
We define the state-independent pairing gap as follows:
\begin{equation}
\label{Delta}
\Delta(t) \equiv g\sum_{k>0} \kappa_k(t) f(\epsilon_k^0) .
\end{equation}
The gap parameter for each canonical pair of states,
$k$ and $\bar k$, can be written as
$\Delta_k(t)=\Delta(t) f(\epsilon_k^0)$.

\subsection{Energy density functional and coordinate-space
representation}

In the present calculations,
we adopt a Skyrme energy functional, $E_{\rm Sky}[\rho]$,
with the parameter set of SkM*\cite{Bart82}.
The functional contains both time-even and time-odd densities
same as Ref. \cite{BFH87}.
The pairing energy functional is added to this,
to give the total energy functional,
$E[\rho,\kappa]=E_{\rm Sky}[\rho]+E_g[\kappa]$.

We use the Cartesian coordinate-space representation for
the canonical states,
$\phi_k(\vec{r},\sigma;t)=\inproduct{\vec{r},\sigma}{\phi_k(t)}$
with $\sigma=\pm 1/2$.
The three-dimensional (3D) coordinate space is discretized in square mesh of
$\Delta x=\Delta y=\Delta z=0.8$ fm in a sphere with radius of 12 fm.
Thus, each canonical state is represented by
$\phi_k(i,j,k,\sigma;t)$ with three discrete indexes for the 3D space.

\subsection{Calculation for the ground state}

First, we need to obtain a static solution to construct an initial state
for the time-dependent calculation.
The numerical procedure is as follows:
\begin{enumerate}
\item Solve Eq. (\ref{phi_0}) for occupied canonical states ($|k|\leq N/2$)
      with $\rho_k=1$, and construct the HF Hamiltonian $h_0[\rho]$,
      using the imaginary-time method\cite{DFKW80}.
\item Calculate unoccupied canonical states
      $\phi_k^0(\vec{r},\sigma)$ ($|k|>N/2$)
      below the energy cut-off $e_{\rm c}$,
      using the imaginary-time method with $h_0$.
\item Solve the BCS equations\cite{RS80} to obtain $\rho_k$ and $\kappa_k$.
\item Update $h_0[\rho]$ with new $\rho_k$, then solve Eq. (\ref{phi_0})
      again with the imaginary-time method,
      to calculate canonical states with $\epsilon_k^0<e_{\rm c}$.
\item Back to 3. and repeat until convergence.
\end{enumerate}
To solve Eq. (\ref{phi_0}), the imaginary-time-evolution operator
for a small time
interval $\Delta t$ is repeatedly operated on each single-particle wave
function.
After each evolution, the single-particle wave functions are orthonormalized
with the Gram-Schmidt method from low- to high-energy states.
We add the constraints for 
the center of mass,
$\int \vec{r} \rho(\vec{r})=0$,
and the principal axis,
$\int r_i r_j \rho(\vec{r})=0$ ($i\neq j$)
for deformed nuclei.

\subsection{Real-time calculation for strength functions}

The canonical states in the ground state
define the initial state for time evolution.
In order to study the linear response,
we use an weak instantaneous external field of
$V_{\rm ext}(\vec{r},t)=-\eta F(\vec{r}) \delta(t)$ to
start the time evolution.
Here, $F(\vec{r})$ is a one-body field, such as $E1$ operator with
recoil charges,
\begin{equation}
\label{E1}
F_i(\vec{r})=\begin{cases}
  (Ne/A) r_i & \text{for protons} \\
- (Ze/A) r_i & \text{for neutrons}
\end{cases} ,
\end{equation}
where $i=(x,y,z)$.
We also study the isoscalar quadrupole response with
\begin{equation}
F(\vec{r})=\frac{1}{\sqrt{2(1+\delta_{K0})}}
\left\{
r^2Y_{2K}(\hat{\bf r}) +r^2Y_{2-K}(\hat{\bf r})
\right\}
, \quad K=0\mbox{ and } 2 .
\end{equation}
Then, at time $t=0+$, the canonical states are given by
\begin{equation}
\label{phi_k_ini}
\phi_k(\vec{r},\sigma;t=0+) = e^{i\eta F(\vec{r})} \phi_k^0(\vec{r},\sigma) ,
\quad
\phi_{\bar k}(\vec{r},\sigma;t=0+) =
 e^{i\eta F(\vec{r})} \phi_{\bar k}^0(\vec{r},\sigma) , 
\end{equation}
and the BCS factors by
\begin{equation}
\rho_k(t=0+)=\rho_k^0, \quad \kappa_k(t=0+)=\kappa_k^0 .
\end{equation}
The parameter $\eta$ controls the strength of the external field.
In this paper, since we calculate the linear response,
it should be small enough to validate the linearity.

To solve the Cb-TDHFB equations in real time,
we use the simple Euler's algorithm.
\begin{eqnarray}
 i\phi_k(t+2dt) &=& i\phi_k(t) +  \{
h(t+dt)-\epsilon_k(t+dt)\} \phi_k(t+dt) \times 2dt, \\
 i\rho_k(t+2dt) &=& i\rho_k(t)
 +\{ \kappa_k(t+dt) \Delta_k^*(t+dt)- \text{c.c.}\}\times 2dt, \\
 i\kappa_k(t+2dt) &=& i\kappa_k(t)
 +[ \kappa_k(t+dt) \{\epsilon_k(t+dt) +\epsilon_{\bar{k}}(t+dt)
                     -2\lambda \}
 \nonumber \\
 && +\Delta_k(t+dt) \{2\rho_k(t+dt)-1\} ] \times 2dt  .
\end{eqnarray}
Here, we insert the chemical potential in Eq. (\ref{dkappa_dt})
which cancels a global time-dependent phase at the ground state,
$e^{-2i\lambda t}$, for $\kappa_k$ and $\Delta_k$.
To construct the states at the first step of $t=dt$,
we use the fourth-order Taylor expansion of the time-evolution
operator for the canonical states\cite{NY05}
and use the Euler's method for
$\rho_k(dt)$ and $\kappa_k(dt)$.
The time step $dt$ is 0.0005 MeV$^{-1}$.
The time evolution is calculated up to $T=10$ MeV$^{-1}$.

The strength function with respect to the operator $F$
is calculated with the following formula\cite{NY05}.
\begin{equation}
S(E;F) \equiv \sum_{n}|\bra{\Phi_{n}} F \ket{\Phi_{0}}|^{2}\delta (E-\tilde{E}_{n}) = -\frac{1}{\pi\eta}{\rm Im} f(E) ,\quad
\tilde{E}_{n} > 0 \ ,
\end{equation}
where $\tilde{E}_n=E_n-E_0$ and $f(E)$ is defined by
\begin{equation}
f(E)=\int_0^\infty dt \ e^{(iE-\Gamma/2)t} 
\int F(\vec{r}) \left\{ \rho(\vec{r},t) -
                     \rho(\vec{r},0) \right\} d\vec{r} ,
\end{equation}
where we have introduced a smoothing parameter $\Gamma$ which is
set to 1 MeV throughout the calculations in Sec.~\ref{sec:numerical_results}.
The formula can be obtained from the time-dependent perturbation theory
in the first order with respect to $\eta$ \cite{NY05}.
Note that the strength function $S(E;F)$ is independent from
magnitude of the parameter $\eta$ as far as the linear approximation
is valid.
In the present study, we adopt the value of $\eta=10^{-4}$ fm$^{-1}$
for the $E1$ operator, and $\eta=10^{-3}$ fm$^{-2}$
for the quadrupole operator.

\section{Numerical results of linear response calculation}
\label{sec:numerical_results}

In this paper, we apply the Cb-TDHFB method to calculation of the
strength functions for Ne and Mg isotopes.
First, we show, in Table \ref{tab: gs_properties}, calculated
ground-state properties; deformations, chemical potentials,
and gap energies defined
by Eq. (\ref{Delta}).
These nuclei show a variety of shapes (spherical, prolate, oblate,
and triaxial), with and without superfluidity.
For nuclei in the superfluid phase with $\Delta\neq 0$,
the numbers of canonical orbitals, $M_\tau$,
included in the calculation ($e_k^0<e_{\rm c}$) are as follows:
For protons,
$M_p=16$ for $^{24,26,28}$Ne and for $^{26,28,30,32}$Mg, and
$M_p=20$ for $^{30}$Ne.
For neutrons,
$M_n=20$ orbitals for $^{28}$Ne,
$M_n=24$ for $^{32}$Ne,
$M_n=28$ for $^{30,34,36}$Mg,
$M_n=30$ for $^{38,40}$Mg.
These numbers are, of course, larger than the proton and neutron numbers,
but not significantly different.
In the case with $\Delta=0$, of course, we only calculate the occupied
orbitals ($M_p=Z$ and $M_n=N$).
Therefore, the numerical task of the Cb-TDHFB
is in the same order as that of the TDHF.
Note that, in the real-time calculation,
the numerical cost is proportional to $M_n+M_p$.

\begin{table}[ht]
\caption{Calculated ground-state properties of Ne and Mg isotopes;
quadrupole deformation parameters $(\beta,\gamma)$,
pairing gaps (\ref{Delta}) for neutrons and protons $(\Delta_n,\Delta_p)$,
chemical potentials for neutrons and protons $(\lambda_n,\lambda_p)$.
In the case of normal phase ($\Delta=0$),
we define the chemical potential as the
single-particle energy of the highest occupied orbital,
$\lambda_n=\epsilon_N^0$ and $\lambda_p=\epsilon_Z^0$.
The pairing gaps and chemical potentials are given in units of MeV.}
\label{tab: gs_properties}
\begin{center}
\begin{tabular}{c|lcllrr} \hline\hline
  & $\beta$ & $\gamma$ & $\Delta_n$ & $\Delta_p$ & $-\lambda_n$ & $-\lambda_p$
\\ \hline
 $^{20}$Ne   & 0.37 & 0$^\circ$  & 0.0  & 0.0  & 13.07 & 9.19 \\
 $^{22}$Ne   & 0.37 & 0$^\circ$  & 0.0  & 0.0  & 11.03 & 12.38 \\
 $^{24}$Ne   & 0.17 & 60$^\circ$ & 0.0  & 0.74 & 10.57 & 13.04 \\
 $^{26}$Ne   & 0.0  & $-$        & 0.0  & 1.00 &  7.17 & 14.92 \\
 $^{28}$Ne   & 0.0  & $-$        & 0.79 & 1.01 &  3.22 & 17.05 \\
 $^{30}$Ne   & 0.0  & $-$        & 0.0  & 1.01 &  3.79 & 19.09 \\
 $^{32}$Ne   & 0.36 & 0$^\circ$  & 0.95 & 0.0  &  2.16 & 23.61 \\
\hline
  $^{24}$Mg & 0.39 & 0$^\circ$  & 0.0  & 0.0  & 14.12 & 9.51 \\
  $^{26}$Mg  & 0.20 & 54$^\circ$ & 0.0  & 0.86 & 13.08 & 11.23 \\
  $^{28}$Mg  & 0.0  & $-$        & 0.0  & 1.03 &  9.21 & 13.30 \\
  $^{30}$Mg  & 0.0  & $-$        & 1.31 & 1.03 &  5.48 & 15.49 \\
  $^{32}$Mg  & 0.0  & $-$        & 0.0  & 1.03 &  5.83 & 17.55 \\
  $^{34}$Mg  & 0.37 & 0$^\circ$  & 1.45 & 0.0  &  4.12 & 20.18 \\
  $^{36}$Mg  & 0.33 & 0$^\circ$  & 1.43 & 0.0  &  3.21 & 21.95 \\
  $^{38}$Mg  & 0.30 & 0$^\circ$  & 1.47 & 0.0  &  2.38 & 23.69 \\
  $^{40}$Mg  & 0.29 & 0$^\circ$  & 0.91 & 0.0  &  1.31 & 25.28 \\
\hline
\end{tabular}
\end{center}
\end{table}

\subsection{Isoscalar quadrupole excitations: Comparison with
QRPA calculations}

We expect that the strength functions calculated in the present real-time
approaches reproduce those in the QRPA.
This is strictly true if we solve the full TDHFB equations, however, since
we solve the Cb-TDHFB equations with the schematic pairing functional
of Eq. (\ref{E_G}) with (\ref{gff}), let us first show the comparison
between our result and the HFB+QRPA calculations.
The QRPA calculations have been done with
a computer program for axially deformed nuclei developed
in Ref. \cite{YG08-2},
which diagonalizes the QRPA matrix of large dimensions in
the quasi-particle basis.
This is based on the HFB ground state calculated in the two-dimensional
coordinate-space representation with the Skyrme
functional SkM* but with the density-dependent contact
interaction for the pairing channel.
The space is truncated by the quasi-particle energy
cut-off of $E_{\rm cut}=60$ MeV and also by the cut-off for
the magnetic quantum
number of the quasi-particle angular momentum, $\Omega_{\rm c}=19/2$.
In this QRPA calculation, the residual spin-orbit
and Coulomb interactions are neglected.
Thus, in order to make a comparison more meaningful,
we also neglect the time-dependence
of these potentials in the Hamiltonian $h(t)$, during the time evolution.

\begin{figure}[t]
\centerline{
\includegraphics[height=0.7\textwidth, angle=-90]{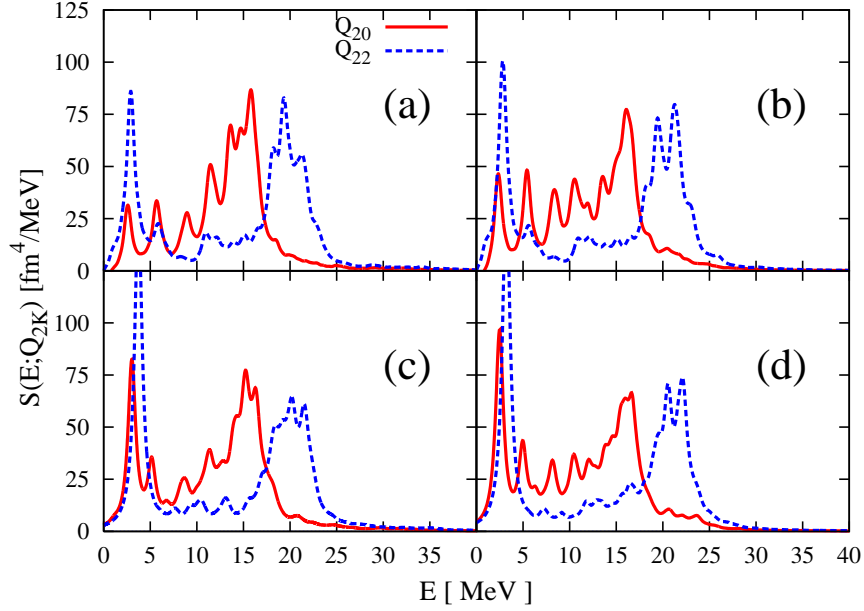}
}
\caption{Calculated isoscalar quadrupole strength distribution for $^{34}$Mg:
(a) Cb-TDHFB with time-independent spin-orbit and Coulomb potentials,
(b) Cb-TDHFB,
(c) Deformed QRPA
without the residual spin-orbit and Coulomb interactions\cite{YG08-2}.
and (d) Deformed QRPA calculation\cite{Los10}.
The smoothing parameter of $\Gamma=1$ MeV is used.
}
\label{fig: ISQ_Mg34}
\end{figure}

In panels (a) and (c) of Fig. \ref{fig: ISQ_Mg34},
we show the isoscalar quadrupole strength
distributions ($K=0$ and 2) for $^{34}$Mg, calculated with
(a) Cb-TDHFB and (c) QRPA.
The ground state has an axially symmetric prolate shape with
finite pairing gaps for neutrons (Table~\ref{tab: gs_properties}).
The HFB calculation with the contact pairing interaction for
the panel (c) produces
a deformation and average neutron pairing gap of
$\beta=0.37$ and $\Delta_n = 1.7$ MeV, respectively.
Note that a renormalization factor, which was used in Ref. \cite{YG08-2},
is set to be unity in the present QRPA calculation.
The peak energies in these calculations are approximately identical,
however, the height of the lowest peak is noticeably different.
We suppose that this is due to the difference in the pairing energy functionals.

In panels (b) and (d) of Fig. \ref{fig: ISQ_Mg34},
we show another comparison between
the Cb-TDHFB calculation and the QRPA calculation of Losa et al\cite{Los10}
using the transformed harmonic oscillator basis.
Since this QRPA calculation includes all the residual interactions,
it is compared with the Cb-TDHFB calculation with the fully
self-consistent time dependence.
It turns out that the residual spin-orbit and Coulomb interactions
slightly shift the giant quadrupole resonance higher in energy,
while they shift the lowest peak lower in energy.
Actually, these shifts are mainly attributed to the residual spin-orbit
interaction and the effect of the residual Coulomb is very small.
The results in panels (b) and (d) well agree with each other, except for
the height of the lowest peak.
Again, this may be due to the different treatment of the pairing,
because Ref. \cite{Los10} also uses the contact pairing interaction.
These comparisons indicate that the small-amplitude Cb-TDHFB calculation
well reproduces a fully self-consistent QRPA calculations.
We would like to mention that, for the isovector dipole excitations,
the agreement is even better than the isoscalar quadrupole cases.

\subsection{Isovector ($E1$) dipole excitations}

\begin{wrapfigure}{l}{0.4\textwidth}
\includegraphics[height=0.4\textwidth, angle=-90]{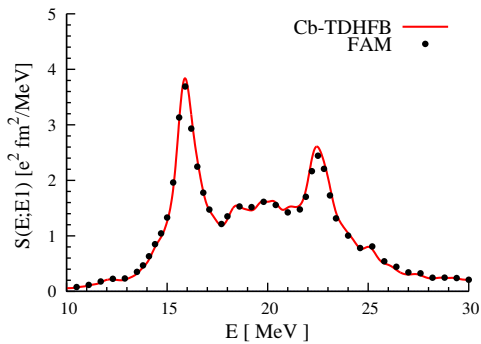}
\caption{$E1$ strength distribution for $^{24}$Mg
calculated with the Cb-TDHFB (solid line) and with
the FAM\cite{NIY07,INY09} (symbols).
The smoothing parameter of $\Gamma=1$ MeV is used.
}
\label{fig: Mg24_FAM}
\end{wrapfigure}
Here, we discuss properties of the isovector dipole excitations including
low-energy pygmy dipole resonances (PDR) and
high-energy giant dipole resonances (GDR).
First, let us show in Fig. \ref{fig: Mg24_FAM} that the comparison
between results of the present Cb-TDHFB calculation and those of
the RPA calculation.
The fully self-consistent RPA calculation has been done with the finite
amplitude method (FAM) developed in Refs.~\cite{NIY07,INY09}.
The same Skyrme functional (SkM*) and the same model space have been
used in these calculations.
Since the ground state of the $^{24}$Mg nucleus is in the normal phase
($\Delta=0$),
these two results should be identical.
This can be confirmed in Fig. \ref{fig: Mg24_FAM},
which clearly demonstrates the accuracy of our real-time method.

\begin{figure}[b]
\includegraphics[height=\textwidth, angle=-90]{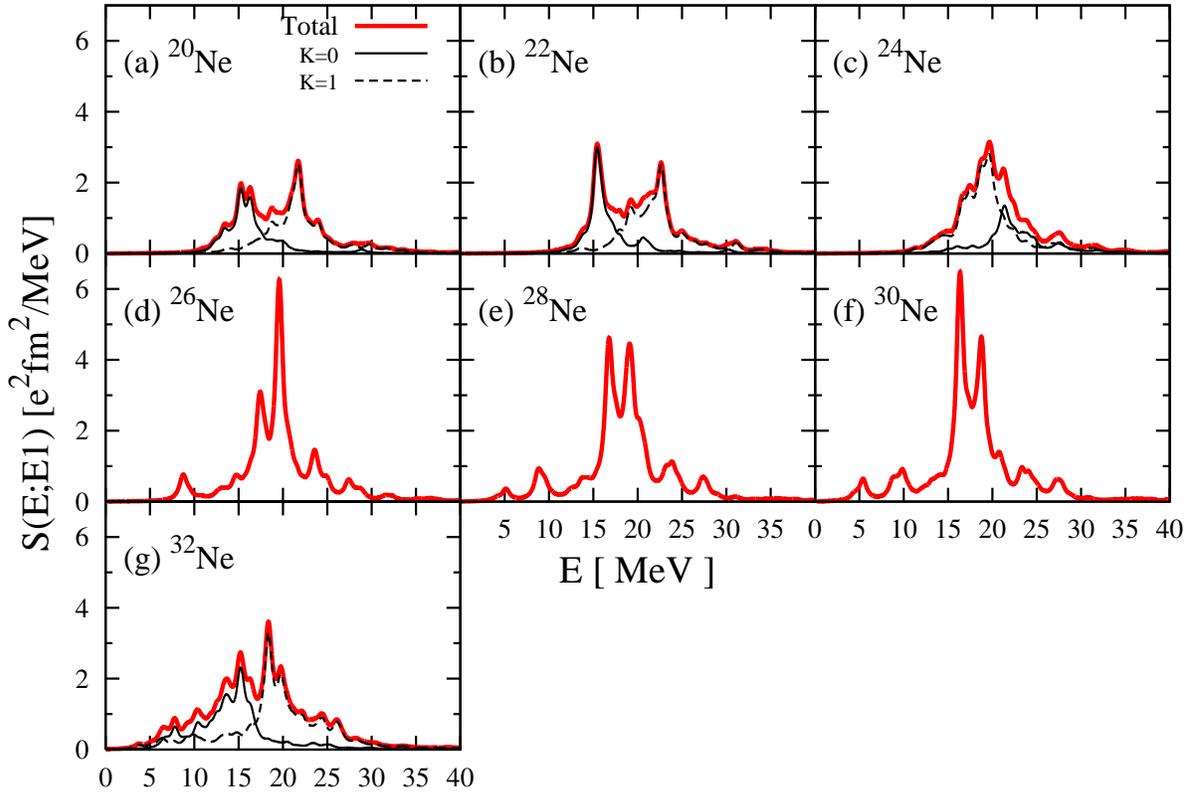}
\caption{Calculated $E1$ strength distribution for Ne isotopes.
For deformed nuclei, the total strength (\ref{S_E1}) is decomposed 
into $S_z(E;(E1))$ (thin solid line) and $S_x(E;(E1))+S_y(E;(E1))$
(dashed line).
The $z$-axis is the symmetry axis for axially deformed cases.
The smoothing parameter of $\Gamma=1$ MeV is used.
}
\label{fig: IVD_Ne}
\end{figure}

Next,
in Fig. \ref{fig: IVD_Ne}, we show calculated $E1$ strength distribution
for Ne isotopes.
Here, $S(E;E1)$ is defined as
\begin{equation}
\label{S_E1}
S(E;E1)=\sum_{i=x,y,z} S_i(E;E1)
=\sum_{i=x,y,z} \sum_n |\bra{n} F_i \ket{0} |^2
            \delta(E-\tilde{E}_n) ,
\end{equation}
where the one-body operator $F_i$ is given by Eq. (\ref{E1}).
The $K=0$ strength is $S_z(E;E1)$
and $K=1$ corresponds to $S_x(E;E1)+S_y(E;E1)$.
Here, for axially symmetric nuclei, the symmetry axis is chosen as $z$-axis.
The ground states of $^{20,22}$Ne are deformed in a prolate
shape with $\Delta=0$ for both protons and neutrons.
Thus, the calculation is identical to the small-amplitude TDHF.
Both nuclei show a prominent double peak structure.
The lower peak is located around 16 MeV and the higher one around 22 MeV.
This comes from the deformation splitting, and
the lower peak is characterized as $K=0$ and the higher one as $K=1$.
The similar structure is seen in the neutron-rich nucleus $^{32}$Ne.
However, despite of the fact that the magnitude of deformation is
roughly same as that of $^{20,22}$Ne, the position of
the higher peak ($K=1$) is lowered and the splitting is not as
prominent as that in $^{20,22}$Ne.
In oblate nuclei such as $^{24}$Ne,
the deformation splitting is not clearly seen in the total strength
distribution, $S(E;E1)$, because the high-energy peak becomes much smaller
than the lower peak.

For $^{24-32}$Ne, calculated ground states are in the superfluid phase
for either neutrons or protons or both.
Peak energies of the giant dipole resonance (GDR) gradually
decrease as the neutron number increases, from about 20 MeV to 17 MeV.
We have confirmed that the pairing correlation does not significantly
affect the $E1$ strength distribution.
However, for some cases, the ground-state deformation is changed by
the presence of the pairing.
For instance, 
the $^{26}$Ne nucleus is deformed in the prolate shape if we neglect the
pairing correlations.
In contrast, the present BCS calculation produces the spherical ground
state.

The low-energy $E1$ strength, which is often called ``pygmy resonance'',
is of significant interests.
In Ne isotopes, there are two effects to create the low-energy $E1$ strength:
One is a large deformation splitting which
brings the lower peak down to around 15 MeV.
Another effect comes from the neutron excess.
In $^{26-32}$Ne, the pygmy peaks appear below 10 MeV.
For $^{26}$Ne, this low-energy peak structure has been recently measured
at RIKEN\cite{Gib08}.
The calculated pygmy position is around $8-9$ MeV, which
agrees with experimental data\cite{Gib08} and
with the other QRPA calculations\cite{CM05,YG08-2}.
For nuclei with even more neutrons ($A \geq 28$), there appear
a double-peak structure below 10 MeV.

\begin{figure}[tb]
\includegraphics[height=\textwidth, angle=-90]{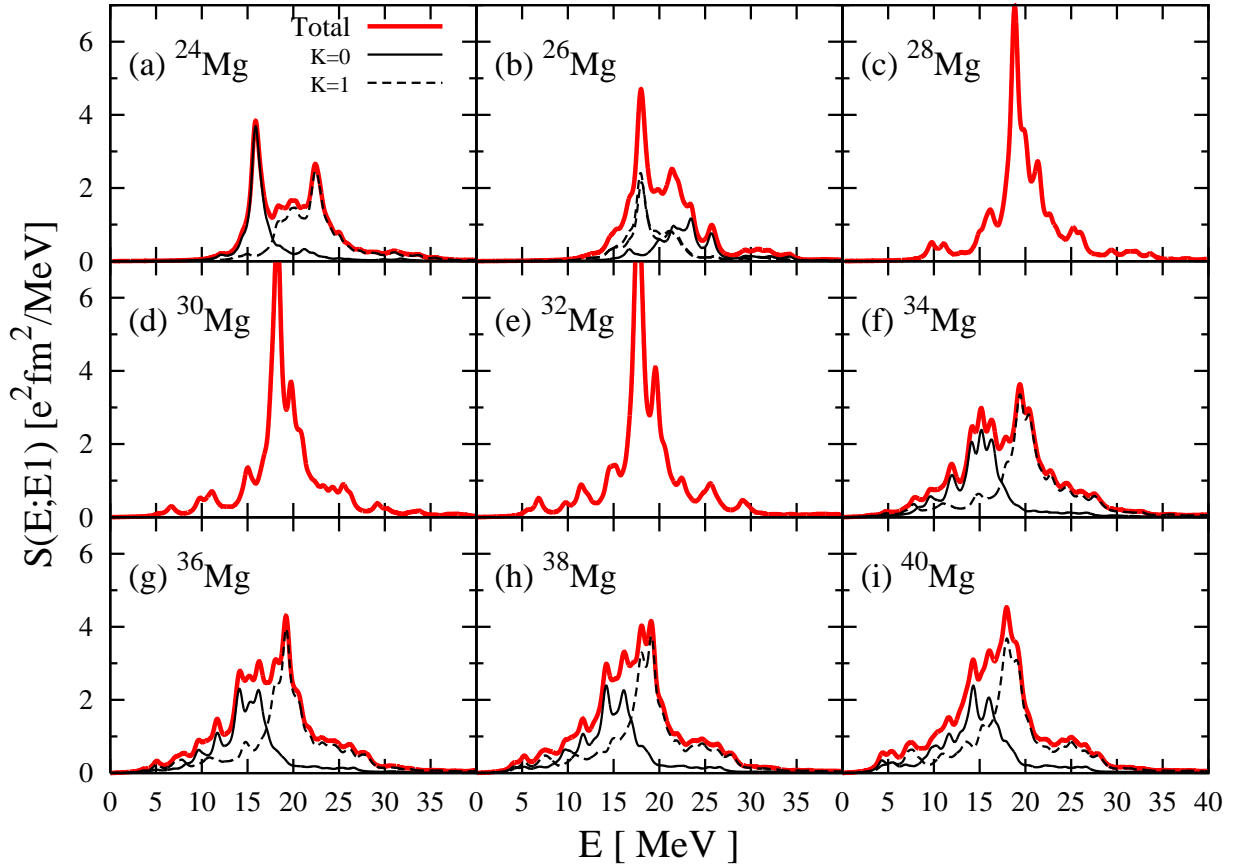}
\caption{Same as Fig.~\ref{fig: IVD_Ne} but for Mg isotopes.
}
\label{fig: IVD_Mg}
\end{figure}

In Fig. \ref{fig: IVD_Mg}, $E1$ strength distribution for Mg isotopes
are displayed.
$^{26}$Mg is nearly oblate, but has a triaxial shape with
$\gamma54^\circ$.
The low energy peak at 18 MeV is prominent in this nucleus.
In $^{28}$Mg and heavier isotopes, there are pygmy states below 10 MeV.
As is in Ne isotopes, there appear a double-peak structure for
$A\geq 30$, though the heights of these pygmy peaks in Mg are lower
than those in Ne isotopes.

\begin{wrapfigure}{r}{0.45\textwidth}
\includegraphics[height=0.45\textwidth, angle=-90]{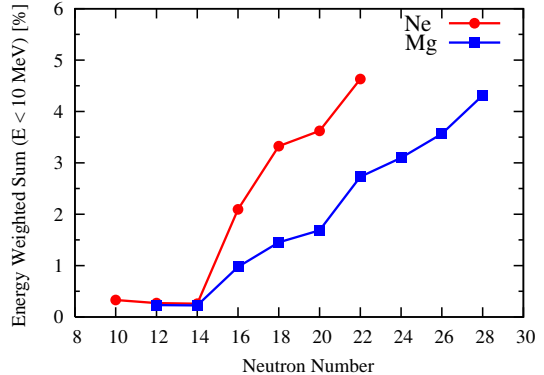}
\caption{Ratio of low-energy $E1$ energy-weighted sum value
to the total sum value, as a function of neutron number.
See text for details.
}
\label{fig: Pygmy}
\end{wrapfigure}
In order to investigate how the low-energy pygmy strength changes as the
neutron number increases,
we define the low-energy $E1$ ratio by $m_1(E_c)/m_1$ with
$E_c=10$ MeV, where
\begin{equation}
m_1(E)\equiv \int_0^E E' S(E';E1) dE' ,
\end{equation}
and $m_1\equiv m_1(\infty)$.
This ratio is shown for calculated even-even Ne and Mg
isotopes in Fig. \ref{fig: Pygmy}.
Both isotopes with $N=10\sim 14$ have very little $E1$ strength below
10 MeV.
Then, the ratio, $m_1(E_c)/m_1$,
shows a rapid increase as a function of neutron number.
Especially, we can see abrupt jumps between $N=14$ and 16, and
between $N=20$ and 22.
The first jump between $N=14$ and 16 seems to be due to
occupation of neutron $s_{1/2}$ orbital.
In contrast, the second jump between $N=20$ and 22 may be due to
the onset of the deformation and the neutron pairing.
The low-energy strengths in Ne isotopes show roughly twice larger 
values compared with Mg nuclei with the same neutron numbers.
This may be attributed to the difference in the separation energy
(chemical potential).

\begin{wrapfigure}{l}{0.5\textwidth}
\includegraphics[height=0.5\textwidth, angle=-90]{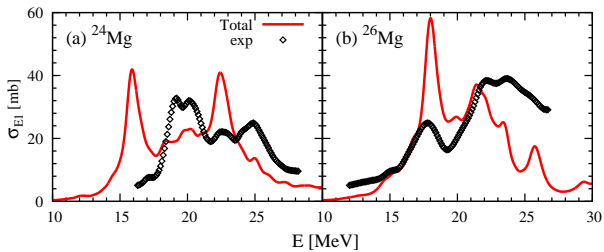}
\caption{Photoabsorption cross sections for $^{24,26}$Mg.
Experimental data (symbols) are taken from Ref. \cite{Varlamov03}.
The smoothing parameter of $\Gamma=1$ MeV is used for the calculations.
}
\label{fig: Mg24_exp}
\end{wrapfigure}
Finally, let us present photoabsorption cross sections in the GDR energy
region ($E=10\sim 30$ MeV),
together with experimental data\cite{Varlamov03}.
For $^{24}$Mg,
the peak energies of the GDR are underestimated by about 3 MeV.
This disagreement has been already found in Ref.\cite{INY09} for $^{24}$Mg.
The present calculation also indicates that this underestimation of the GDR
peak energy is also true for $^{26}$Mg.
The $E1$ strength distribution for $^{26}$Mg is very similar to
that in Fig. 12 (bottom panel) in Ref.~\cite{Los10}.
In light nuclei, the GDR energy is systematically underestimated in
most of the Skyrme functionals\cite{INY09}, that seems to be true
for nuclei with superfluidity.

\section{Conclusion}
\label{sec:summary}

We have developed an approximate approach to the time-dependent
Hartree-Fock-Bogoliubov (TDHFB) theory, using the
canonical-basis representation for time evolution of the
densities and pairing tensors.
Although, in general, the pair potential is not in a diagonal
form in the canonical basis, if it is approximated in such a form,
the TDHFB equations can be enormously simplified, to give
a canonical-basis TDHFB (Cb-TDHFB) equations.
In this paper, we have treated a full Skyrme functional
for the particle-hole channel, however, used a simple schematic
functional for the pairing channel.
Since the schematic pairing functional violates the gauge invariance,
it requires a special choice for the gauge condition.
The Cb-TDHFB equations contain the TDHF as a special case of
absence of pairing correlations.
Its static limit is identical to the HF+BCS approximation.
We have shown that the equations possess many of
desired properties analogous to the original TDHFB theory, including
the average particle number and the average total energy.
We have also investigated analytical properties of its small-amplitude
limit and found that the Nambu-Goldstone modes correspond to zero-energy
normal modes and are automatically separated from other finite-energy modes.

We have developed a computational program for real-time propagation
based on the Cb-TDHFB equations using the three-dimensional (3D)
coordinate-space representation.
To test the accuracy and validity of the present method,
we have calculated the isoscalar quadrupole strength distribution
in deformed $^{34}$Mg
with the small-amplitude real-time method and compared with
deformed QRPA calculations with a standard diagonalization
method\cite{YG08-2,Los10}.
Results well agree with each other, except for the quadrupole strength
of the lowest state located around 3 MeV.
This may be due to the difference of the pairing energy functional
used in the Cb-TDHFB and QRPA calculations.

Then, we have calculated the $E1$ strength distribution for
even-even Ne and Mg isotopes systematically.
The ground-state properties of these isotopes changes from one
nucleus to another.
For instance, there are a variety of shapes including spherical,
prolate, oblate, and triaxial deformations.
The gap energies also significantly changes, depending on
the particle number and deformation.
The 3D representation allows us to treat all of these nuclei
in a self-consistent and systematic manner.
Typical deformation splitting of the giant dipole resonances (GDR) is
predicted for prolate deformed nuclei, $^{20,22,32}$Ne and $^{24,34-40}$Mg.
The neutron-rich deformed nuclei, such as $^{32}$Ne and $^{34-40}$Mg,
show a $K=0$ peak around 15 MeV and a significant strength in a
low-energy tail at $5\sim 10$ MeV.
The low-energy $E1$ pygmy strength is almost negligible for
$^{20-24}$Ne and $^{24-26}$Mg, but suddenly starts increasing at the
neutron number 16 and another jump at 22.
This seems to be due to the occupation of the neutron $s_{1/2}$ orbital
and the onset of the neutron pairing.
The effect of the deformation also plays a role for the increase of the
pygmy strength in low-energy region.
These low-energy $E1$ strength is of significant interest in studies
of the element-synthesis reactions in stars and
in explosive environments.

The Cb-TDHFB method is easily applicable to heavier systems.
Its computational task is roughly same as that of the TDHF.
Furthermore, adopting the pairing functional calculated from
an interaction instead of the schematic one,
it can be used for calculation of the heavy-ion collision dynamics
beyond the TDHF, including dissipative dynamics induced by the
pairing interaction.

\appendix*
\section{Proof of properties of Cb-TDHFB with $E_g$}

Among the properties listed in Sec.~\ref{sec: properties_E_G},
the conservation of the orthonormal property and
that of the particle number are trivially identical to
Sec. \ref{sec: properties}.
In the followings, we show a simple proof of the other properties.

\subsection{Average total energy conservation}

Using the Cb-TDHFB equations, it is easy to show that
the time derivative of the schematic pairing functional of Eq. (\ref{E_G}),
gives 
\begin{equation}
i\frac{d}{dt}E_g=-i\sum_{k,l>0} G_{kl} \left(
\frac{d\kappa_k^*}{dt} \kappa_l(t)
+\kappa_k^*(t) \frac{d\kappa_l}{dt} \right)
=-i\sum_{k>0} \frac{d\rho_k}{dt}
(\eta_k(t) + \eta_{\bar k}(t)) .
\end{equation}
Only with the special choice of the gauge parameters,
(\ref{gauge_fix}),
we observe the conservation of the total energy.

\subsection{Stationary solution}

Following the arguments in Sec.~\ref{sec: static_limit},
it is easy to see that
the stationary solution corresponds to the ordinary HF+BCS result,
but only when we adopt the gauge fixing (\ref{gauge_fix}).

\subsection{Small amplitude limit}

With use of the pairing functional (\ref{E_G}),
we can no longer assume the time-dependent phase factor of
Eq. (\ref{Delta_phase}) for $\Delta_k(t)$.
Instead, we only extract the global phase related to the chemical potential
from $\kappa_k(t)$ and $\Delta_k(t)$.
\begin{eqnarray}
\ket{\phi_k(t)}=\ket{\phi_k^0} + \ket{\delta\phi_k(t)} , \quad
\ket{\phi_{\bar k}(t)}=\ket{\phi_{\bar k}^0} + \ket{\delta\phi_{\bar k}(t)} , \\
\kappa_k(t)=\{ \kappa_k^0 + \delta\kappa_k(t)\}e^{-2i\lambda t} , \quad
\Delta_k(t)=\{ \Delta_k^0 + \delta\Delta_k(t)\}e^{-2i\lambda t} ,\\
\rho_k(t)=\rho_k^0 + \delta\rho_k(t) , \quad
h(t)=h_0 + \delta h(t) .
\end{eqnarray}
Using the gauge condition (\ref{gauge_fix}),
we have the following equations for the small-amplitude limit of the
Cb-TDHFB equations.
\begin{eqnarray}
\label{ddeltaphi_dt_2}
i\frac{\partial}{\partial t} \ket{\delta\phi_k(t)}
&=& (h_0 - \epsilon_k^0) \ket{\delta\phi_k(t)} +
(1-\ket{\phi_k^0}\bra{\phi_k^0})\delta h(t)\ket{\phi_k^0} ,
\quad (k\leftrightarrow \bar k) , \\
\label{ddeltarho_dt_2}
i\frac{\partial}{\partial t} \delta\rho_k(t) &=&
\Delta_k^0 \delta\kappa_k(t) + \kappa_k^0 \sum_{l>0} G_{kl} \delta\kappa_l(t)
 - \mbox{ c.c.} ,\\
\label{ddeltakappa_dt_2}
i\frac{\partial}{\partial t} \delta\kappa_k(t) &=&
2(\epsilon_k^0-\lambda)\delta\kappa_k(t)
+(\bra{\phi_k^0}\delta h(t) \ket{\phi_k^0}
  + \bra{\phi_{\bar k}^0}\delta h(t) \ket{\phi_{\bar k}^0}) \kappa_k^0
\nonumber \\
&& +(2\rho_k^0-1) \sum_{l>0} G_{kl} \delta\kappa_l(t)
+2\Delta_k^0\delta\rho_k(t) .
\end{eqnarray}
Here, we used the equation,
$\bra{\phi_k^0+\delta\phi_k(t)}(h_0+\delta h(t))\ket{\phi_k^0+\delta\phi_k(t)}
 = \epsilon_k^0+\bra{\phi_k^0}\delta h(t)\ket{\phi_k^0}$,
which can be verified because of the norm conservation.

\subsubsection{Translation and rotation}

The same argument as that of Sec. \ref{sec: translation_rotation}
leads to
\begin{equation}
\label{NG_2}
i\alpha (h_0 - \epsilon_k^0) S \ket{\phi_k^0} + (1-\ket{\phi_k^0}\bra{\phi_k^0})
i \alpha [ S, h_0 ] \ket{\phi_k^0} = 0 ,
\end{equation}
where we multiply the projection $(1-\ket{\phi_k^0}\bra{\phi_k^0})$ on
both sides of Eq. (\ref{NG_1}).
Equation (\ref{NG_2}) means that
$\ket{\delta\phi_k^{S'}}\equiv i (1-\ket{\phi_k^0}\bra{\phi_k^0})
S \ket{\phi_k^0}$ and
$\delta h_S\equiv i[S,h_0]$
correspond to a zero-energy normal-mode solution
for Eq. (\ref{ddeltaphi_dt_2}).
Note that $\ket{\delta\phi_k^S}=i\alpha S \ket{\phi_k^0}$
and $\ket{\delta\phi_k^{S'}}$ produce the identical density fluctuation.
$\delta\rho_k=0$ and $\delta\kappa_k=0$ also satisfy Eqs. (\ref{ddeltarho_dt_2})
and (\ref{ddeltakappa_dt_2}),
since $\bra{\phi_k^0} \delta h_S \ket{\phi_k^0}
=i\bra{\phi_k^0} [ S, h_0 ] \ket{\phi_k^0}=0$.
Therefore,
the translational and rotational modes appear as the zero-energy normal modes.

\subsubsection{Pairing rotation}

When the ground state is in the superfluid phase,
the transformation $e^{i\theta N}$ 
changes the phase of $\kappa_k$.
\begin{eqnarray}
\delta\kappa_k^N &=& e^{2i\theta} \kappa_k^0 -\kappa_k^0\approx
2i\theta\kappa_k^0, \\
\delta\rho_k^N &=& \delta h_N = 0, \quad
\ket{\delta\phi_k^N} = \ket{\delta\phi_{\bar k}^N} = 0 .
\end{eqnarray}
Using Eq. (\ref{kappa_0}), it is easy to see that
these quantities correspond to a normal-mode solution with $\omega=0$ for
Eqs. (\ref{ddeltaphi_dt_2}), (\ref{ddeltarho_dt_2}),
and (\ref{ddeltakappa_dt_2}).
Thus, the pairing rotational modes 
appear as the zero-energy modes.
Again, without the gauge condition (\ref{gauge_fix}),
we would not have this property.

\subsubsection{Particle-particle (hole-hole) RPA}

In case of $\kappa_k^0=0$,
it is easy to see that the p-p (h-h) channels are
decoupled from the p-h channels.
The p-p and h-h dynamics are described by the following equations.
\begin{equation}
\label{pair_vib_2}
i\frac{\partial}{\partial t} \delta\kappa_k(t) =
2(\epsilon_k^0-\lambda)\delta\kappa_k(t)
\pm \sum_{l>0} G_{kl} \delta\kappa_l(t) ,
\end{equation}
where the sign $+$ ($-$) is for hole (particle) orbitals.
Again, introducing the forward and backward amplitudes in the same
way as Sec. \ref{sec: pairing_vibration},
Eq. (\ref{pair_vib_2}) can be rewritten in a matrix form as
\begin{equation}
\begin{pmatrix}
2\epsilon_p^0\delta_{pp'} -G_{pp'} & G_{ph'} \\
G_{hp'} & -2\epsilon_h^0\delta_{hh'} -G_{hh'}
\end{pmatrix}
\begin{pmatrix}
Z_{p'} \\
Z_{h'}
\end{pmatrix}
= \omega
\begin{pmatrix}
1 & 0 \\
0 & -1 \\
\end{pmatrix}
\begin{pmatrix}
Z_p \\
Z_h
\end{pmatrix} ,
\end{equation}
where
$Z_p= X_p$ ($Z_p= Y_p$) and $Z_h=Y_h$ ($Z_h=X_h$) for the p-p (h-h) channel.

\begin{acknowledgments}
This work is supported by Grant-in-Aid for Scientific Research(B)
(No. 21340073) and on Innovative Areas (No. 20105003).
Computational resources were provided by the PACS-CS project and
the Joint Research Program
(07b-7, 08a-8, 08b-24, 09b-9, 09a-25, 10a-22)
at Center for Computational Sciences,
University of Tsukuba, and by the Large Scale Simulation Program
(Nos. 07-20, 08-14, 09-16, 09/10-10) of High Energy Accelerator Research
Organization (KEK).
Computations were also performed on the RIKEN Integrated Cluster of Clusters
(RICC).
We would like to thank the JSPS Core-to-Core Program ``International
Research Network for Exotic Femto Systems'' and the UNEDF SciDAC
collaboration under DOE grant DE-FC02-07ER41457.
\end{acknowledgments}

\bibliography{nuclear_physics,myself,current}

\end{document}